\documentclass[aps,prl,preprint,groupedaddress]{revtex4-1}

\usepackage{graphicx} 

\begin{document}


\title{A fluid bilayer phase in aqueous mixtures of  fatty alcohol and cationic surfactant}



\author{Tiago Espinosa de Oliveira}
\affiliation{Institut Charles Sadron, CNRS and University of Strasbourg, rue du Loess, F-67034 Strasbourg cedex 2, France}
\affiliation{Universidade Federal do Rio Grande do Sul, Porto Alegre, Brazil}
  \author{Fabien Leonforte}
  \author{Luc Nicolas-Morgantini}
    \author{Anne-Laure Fameau}
  \author{Bernard Querleux}
\affiliation{L'Or\'eal Research and Innovation, France.}
\author{Fabrice Thalmann}
\author{Carlos M. Marques}
  \affiliation{Institut Charles Sadron, CNRS and University of Strasbourg, rue du Loess, F-67034 Strasbourg cedex 2, France}

\date{\today}

\begin{abstract}
The $L_\alpha$ phase of lipid bilayers is a fluid self-assembled state, key to the formulation of cosmetics, detergents and pharmaceutics. Despite having been extensively scrutinized in self-assembled phospholipid or surfactant bilayers, the formation of a fluid $L_\alpha$ state has defied understanding in mixtures of fatty alcohols, surfactants and water, where is viewed as the essential step for the preparation of creamy dispersions. Here, atomistic molecular dynamics simulations show the existence of a fluid bilayer in aqueous mixtures of cetyl (C$_{16}$OH) and stearyl (C$_{18}$OH) alcohols, and cetyl-trimethylammonium chloride (CTAC). These simulated bilayer systems display not only a rich temperature phase diagram with many of the features seen in experiments but carry also the unambigous signature of fluid bilayer behavior.
\end{abstract}


\maketitle

Fatty alcohols (FA) are amphiphilic molecules associating a polar hydroxyl head to an alkyl chain tail (${\rm CH_3 - (CH_2)}_{n-1} {\rm - OH}$)~\cite{2000_noweck_grafahrend}. Cetyl ($n=16$) and stearyl ($n=18$) alcohols, in combination with non-ionic or ionic surfactants, are common ingredients in formulations of cosmetic and pharmaceutical cream products~\cite{1968_barry,1976_eccleston,1976_fukushima_yamaguchi,1987_rowe_mcmahon,1990_goetz_el-aasser,2011_awad_olsson}. These aqueous dispersions, often known as lamellar gel networks, derive their advantageous properties from an extended and highly interconnected lamellar structure~\cite{2017_sakamoto_yamashita}. The basic network units are rigid ($L_\beta$) bilayers, or stacks of bilayers, forming a percolating  structure that can withstand elastic deformations~\cite{1987_benton_wells,2000_eccleston_towns-andrews,2013_iwata_aramaki}.

Depending on temperature, fatty alcohols alone adopt various ordered structures with poor hydration capacity~\cite{1958_tanaka_hayashida,1959_tanaka_hayashida,1977_fukushima_harusawa}, preventing mixtures of water and fatty alcohols  to be formulated as creams with the desired homogeneity, stability and viscoelasticity: only when an appropriate proportion of surfactants is added to the mixture can the lamellar gel network structure be obtained~\cite{2017_sakamoto_yamashita}. 

The textural properties of the cream are empirically optimized by various preparation steps involving combinations of stirring and heating or cooling. But in all formulation variants the different components are first brought together at high enough temperatures where a fluid bilayer ($L_\alpha$) phase is believed to form~\cite{2016_wunsch_flick}. Despite being a key determinant for the formation of the lamellar network, the nature of such high temperature phase of fatty alcohols/surfactant mixtures has not been throughly inspected. In particular, to our knowledge there are no published molecular dynamics simulations of mixtures of water, fatty alcohol and surfactants.

In this Letter we perform all-atom molecular dynamics simulations of aqueous solutions of cetyl (C$_{16}$OH) and stearyl (C$_{18}$OH) alcohols and cetyl-trimethylammonium chloride (CTAC), a typical mixture leading to the formation of lamellar gel networks~\cite{2017_sakamoto_yamashita,2000_eccleston_towns-andrews,2016_wunsch_flick}.  Cetyltrimethyl-ammonium chloride (CTAC)  is a cationic surfactant that mixes well with cetyl and stearyl alcohols that have similar chain lengths. Fatty alcohols (FA) and surfactant (CTAC) molecules were parameterized using the CHARMM~36~\cite{2004_mackerell,2010_klauda_pastor} force-field. In particular, we used for the alkyl chains the same parameters as the ones introduced initially for modeling 1,2-dipalmitoyl-phosphatidylcholine (DPPC) tails. As a matter of fact, CHARMM-36 was shown to accurately describe the melting transition of DPPC lipid bilayers, as a result of some empirical optimization. Methanol CH$_{3}$OH,  and DPPC phosphocholine parameters were used respectively to create the alcohol and trimethyl ammonium groups. The TIP3P water model was used for the solvent.

%


Most of the simulated systems were made of 256 chains arranged into a symmetric bilayer conformation, each opposing leaflet containing 128 chains, and solvated with 2560 water molecules. Initial configurations were  prepared by molecular modeling. Four chemical compositions were considered:  pure cetyl (C$_{16}$OH), 20\% molar surfactant (C$_{16}$OH:CTAC (4:1)), equimolar cetyl-stearyl mixtures without (C$_{16:18}$OH) and with 20\% molar  (C$_{16:18}$OH:CTAC (4:1)) surfactant. Triplicate 100~ns runs for annealing and quenching were performed to thermalize the systems~\cite{supplemental}. This choice for initial configurations was justified by a number of available experimental results consistent with a lowly hydrated gel L$_{\beta}$ structure at room temperature~\cite{2016_wunsch_flick}. 

\begin{figure}[h!]
\center
\includegraphics[width=0.42\textwidth,angle=0]{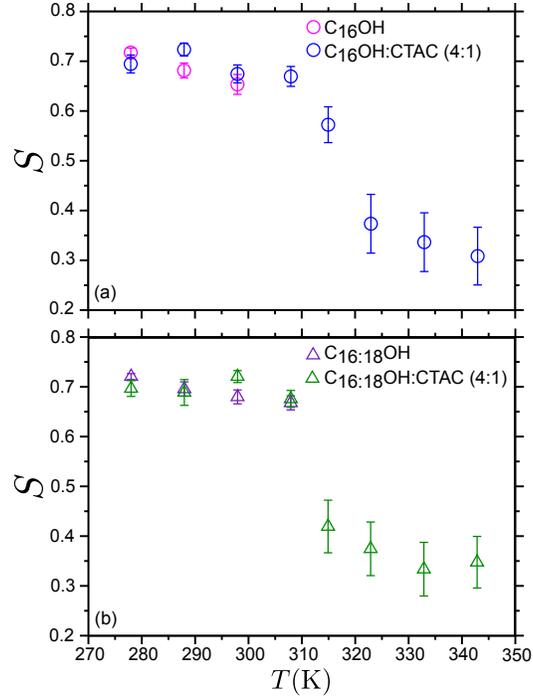}
\caption{Average order parameter as a function of temperature for four different compositions.
(a) Aqueous C$_{16}$OH and C$_{16}$OH:CTAC (4:1) and 
(b) aqueous C$_{16:18}$OH and C$_{16:18}$OH:CTAC (4:1).}
\label{fig:order}
\end{figure}

Structural changes in the FA organisation were followed by computing the carbon-carbon order parameter $S$ of the alkyl chain with respect to the bilayer normal direction $z$. It is defined for every group $(\mathrm{CH}_2)_i$ in a chain as:
\begin{equation}
S_i = \frac{1}{2} \left \langle 3 {\rm cos}^2 \theta - 1 \right \rangle ,
\end{equation}
where $\theta$ is the angle between the two nearest carbons surrounding the selected CH$_2$ group and the membrane normal axis. Bilayer systems are characterized by a positive $S$, which is larger for gel  than for fluid phases. Isotropic structures lead to a vanishing order parameter $S_i=0$. In all the cases of interest, the order parameter decreases as $i$ shifts from the FA head towards the terminal methyl group (-CH$_3$). An average order parameter $S$, representative from the central section of FA chains (averaged over 8 carbons, numbered from 4 to 11), was computed and represented in Fig.~\ref{fig:order}.


%
%


\begin{figure}[h!]
\center
\includegraphics[width=0.45\textwidth,angle=0]{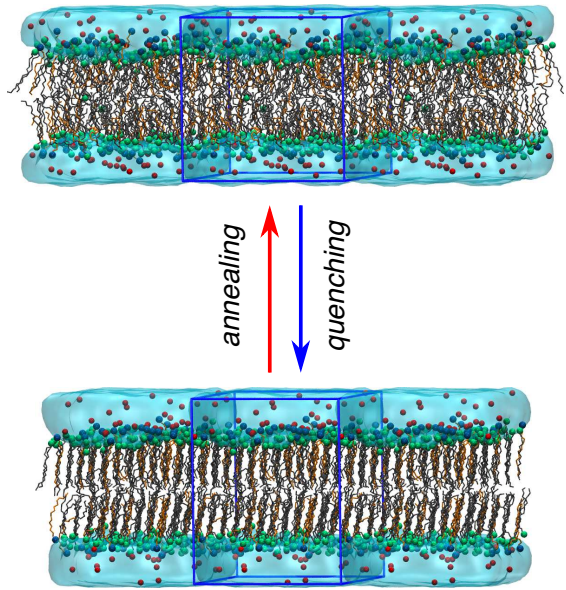}
\caption{Snapshots of the bilayer structures of the  C$_{16:18}$OH:CTAC (4:1) fatty alcool-surfactant mixture at low ($\mathrm{L}_{\beta}$) and high ($\mathrm{L}_{\alpha}$) temperature.}
\label{fig:annealing}
\end{figure}



In the absence of surfactant, both C$_{16}$OH and C$_{16:18}$OH showed a well defined \textit{gel} phase at temperatures below respectively 298 and 308~K, with $S\sim 0.7$. In this phase, alkyl chains adopt well packed dense \textit{all-trans} conformations. At larger temperatures, these two systems were found in a completely disordered phase, presumably an isotropic fluid phase. The preferential arrangement of FA molecules into leaflets was lost. Cetyl-stearyl mixtures melted at a slightly higher temperatures than pure cetyl systems, in agreement with experimental data~\cite{1978_Yamaguchi_Fukushima}.

In the presence of a 1:4 CTAC surfactant ratio (20\% of the system molar mass), a structure consistent with a lamellar fluid phase was observed in a [323-343~K] temperature range. Tails were more disordered, melted, as shown by the lower value $S\sim 0.35$. Experimental systems of comparable FA-CTAC composition but with much larger water contents display a $L_\beta \to L_\alpha$ transition temperature at 243~K.  In our simulations, above 358~K, an isotropic fluid phase, similar to the one seen without CTAC prevailed. We observed that repeated cycles of annealing-quenching resulted in a denser, better packed low temperature solid structure, with an average order parameter higher than 0.7 -- see Fig.~\ref{fig:annealing}. 

\begin{figure}[h!]
\center
\includegraphics[width=0.31\textwidth,angle=0]{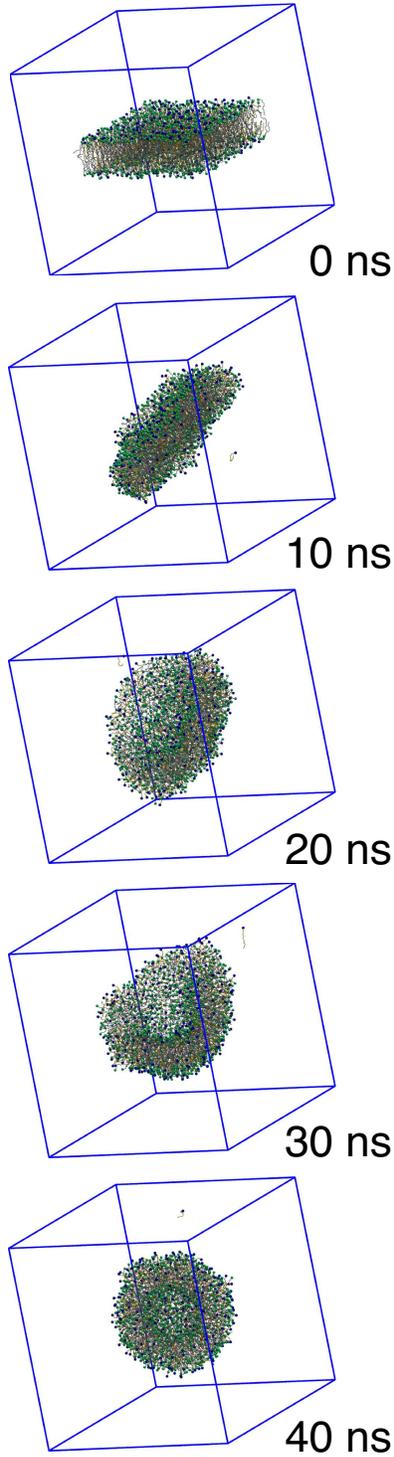}
\caption{Spontaneous formation of a small unilamellar vesicle from a C$_{16:18}$OH:CTAC (4:1) fatty alcool-surfactant mixture at 333 K.}
\label{fig:vesicle}
\end{figure}

As further evidence of a lamellar fluid state stabilized by surfactants, we increased the size of the simulated systems and tested the ability of the bilayer to spontaneously form a vesicle, a topological  hallmark of fluid bilayers. A large system comprising 2560 fatty chains was initially prepared as a flat bilayer. Then, a flat square slab was cut off and dunked into a large water reservoir, in such a way that the slab did not extend across periodic boundary conditions (PBC) and had a free boundary line exposed to the solvent (Fig.~\ref{fig:vesicle} at 0~ns). As the time sequence of snapshots displayed in Fig.~\ref{fig:vesicle} shows, the slab spontaneously closed on itself, adopting a small unilamellar vesicle shape. From 5~ns to 20~ns, the bilayer formed a temporary \textit{bicelle} disk, closed by a folded leaflet rim. This disk-like structure remained for some time, while undergoing severe surface fluctuations. Then, it started to bend spontaneously after 30~ns and ended up forming the closed vesicle after 40~ns of simulation time. The vesicle stayed stable during the following 500~ns of simulation.


The lamellar fluid phase structure was characterized further by measuring the area per fatty chain $A_f$, and the membrane thickness $D_m$. Both parameters were found to evolve smoothly with temperature except for an abrupt change between 315 and 323~K, supporting evidence of a sharp gel-fluid transition (Table~\ref{tab:1}). The bilayer isothermal stretching elastic modulus $K_A$ was estimated based on equilibrium box area fluctuations~\cite{1999_feller_pastor,2004_marrink,2005_denotter,2015_eerden_marrink,2016_guo_thalmann}, according to the relation:
\begin{equation}
K_A = k_BT \frac{ \left\langle A \right\rangle}{\left\langle A^2 \right\rangle - \left\langle A \right\rangle ^2 },
\label{eq:ka}
\end{equation}
where $\left\langle A \right\rangle$ and $\left\langle A^2 \right\rangle$ are the average value and the mean square fluctuation (variance) of the sample monolayer area $A = L_x L_y$, $k_B$ the Boltzmann constant and $T$ the temperature. The area fluctuations arise from coupling the system to a semi-isotropic Parrinello-Rahman barostat~\cite{1980_parrinello_rahman}  and a Nose-Hoover thermostat~\cite{1984_nose,1985_hoover}. The Parrinello-Rahman barostat is assumed to enforce a constant pressure and vanishing surface tension NP$\gamma$T ensemble, \textit{i.e.} the system conformations contribute to the statistical average with a weight proportional to ${\rm exp} \{- (E + P L_z L_x^2)/k_B T\}$, with energy $E$, pressure $P$ and vanishing $\gamma$. This is achieved by allowing the $L_z$  and $L_x = L_y$ lateral box sizes to be independently rescaled. We found a significant drop in $K_A$ at the transition (Table~\ref{tab:1}), consistent with a lamellar fluid L$_{\alpha}$ phase, with $K_A$ values  comparable to those  of lipid bilayers ($\sim$ 250 mN/m)~\cite{2000_nagle_tristram-nagle}. Note that imposing a semi-isotropic barostat is only possible if the simulated system possesses a finite stretching modulus $K_A$. The highest temperature states occurring in the presence and in the absence of CTAC were shown to be unstable under a semi-isotropic barostat, displaying unbound $L_x$-$L_z$ fluctuations, as expected from isotropic fluid phases.  


\begin{table}[h!]
\centering
\begin{tabular}{|c|c|c|c|}
\hline
$T$ (K) & $A_f ({\rm nm^2})$ & $D_m ({\rm nm})$ & $K_A ({\rm mN/m})$  \\ \hline
278     & 0.225              & 4.074            & 2328.394           \\ \hline
288     & 0.227              & 4.080            & 2274.074           \\ \hline
298     & 0.230              & 4.094            & 2146.307           \\ \hline
308     & 0.233              & 4.102            & 1875.433           \\ \hline
315     & 0.228              & 4.202            & 2077.189           \\ \hline
323     & 0.292              & 3.489            & 216.857            \\ \hline
333     & 0.298              & 3.472            & 226.860            \\ \hline
343     & 0.307              & 3.430            & 173.915            \\ \hline
\end{tabular}
\caption{Geometrical parameters for \textit{gel} and \textit{fluid} bilayers of C$_{16}$OH:CTAC (4:1): $A_f$ mean cross-sectional area per chain, $D_m$ bilayer thickness. The relative statistical uncertainty for $A_f$ and $D_m$ is about~0.003.  Bilayer elastic coefficient $K_A$, from a fluctuation analysis argument. The relative statistical uncertainty turns out to be larger, of the order of 20\% to 30\%. All three parameters change sharply between 315 and 323~K.}
\label{tab:1}
\end{table}


Gel-fluid bilayer transitions are commonly believed to be weakly first order~\cite{2007_book_heimburg}. To gain insight into the thermodynamic transition properties, we estimated the change in enthalpy $\Delta H$ upon melting. The enthalpy of the bilayer and water system $H$ was defined as $H = U + PV$~\cite{2005_seo_tsukihashi}, where $U$ is the time averaged total energy, computed from the sum of all the potential energies in the force field and the translation kinetic energy of the atoms, $P$ the target pressure of the barostat, $\gamma=0$ and $V$ the average simulated volume. Bond vibrations are either treated classically, or frozen in the case of hydrogen covalent bonds (methyl, hydroxyl and water groups). Constrained bonds decrease the number of effective independent degrees of freedom and reduce correspondingly the kinetic energy term. Within the limit of the force-field accuracy (\textit{e.g.} truncation of van der Waals interactions, classical or frozen bond vibrations) the discontinuous change in internal energy and enthalpy should account fairly for the variation in cohesion, internal isomerization and hydration energy contributions which are expected to predominantly contribute to the transition. 

\begin{figure}[h!]
\centering
\includegraphics[width=0.46\textwidth,angle=0]{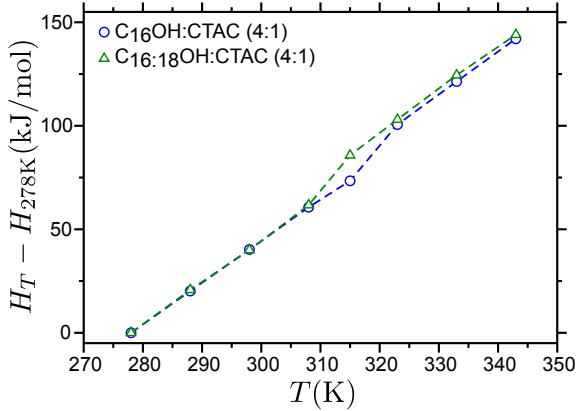}
\caption{Enthalpy as a function of temperature for the C$_{16}$OH:CTAC~(4:1) and C$_{16:18}$OH:CTAC~(4:1) systems.}
\label{fig:H}
\end{figure}

Figure~\ref{fig:H} represents the enthalpy difference $H(T)-H(278~\mathrm{K})$ as a function of temperature for C$_{16}$OH:CTAC~(4:1) and C$_{16:18}$OH:CTAC~(4:1) aqueous solutions. The enthalpy curve shows a smooth trend (constant pressure specific heat) except in the region of the melting transition. The enthalpy variation at the transition of C$_{16}$OH:CTAC~(4:1) and C$_{16:18}$OH:CTAC~(4:1) was estimated by extrapolating the linear trends of the low and high temperature phases. We found respectively for these jumps a value of  10.30 and 10.00 kJ/mol. This result is quantitatively consistent with experimental differential scanning calorimetry data (DSC)~\cite{2016_wunsch_flick}, and falls within the expected range of chain melting values for equivalent phospholipid systems~\cite{2013_marsh_handbook}. We found that the surfactant containing  cetyl and cetyl-stearyl mixtures melts at different temperatures, the longer chain alcohol driving  $L_\beta\to L_\alpha$ melting to higher temperatures,  albeit with comparable transition enthalpies.


To summarize, we have proposed an atomistic model for the cetyl and cetyl-stearyl alcohols mixtures in aqueous solution with and without surfactants (CTAC). We investigated the  structural, mechanical and thermodynamic properties of these systems at various temperatures.  In the absence of surfactant, we found a transition between a solid gel bilayer and an isotropic fluid phase in good agreement with experiments. In the presence of surfactant (20\% CTAC), we found 2 transitions. A first transition was seen between a gel phase and fluid lamellar phase.  Our estimate of the enthalpy of melting is very satisfactory although, perhaps not surprisingly, our temperature transition is lower than that experimentally measured in more hydrated multilamellar systems~\cite{2016_wunsch_flick}. 
We found also that the compressibility modulus falls abruptly upon melting to a value comparable to similar known L$_{\alpha}$ phases. The parameters used in the present work describe well the systems under investigation, both from the thermodynamic and from the mechanic point of view, and we confidently associate the lamellar fluid phase to a L$_{\alpha}$ phase, while the low temperature gel phase is consistent with a L$_{\beta}$ state. Simulation of larger system sizes showed that the fluid bilayer may form stable unilamellar vesicles. 

This work is a first step into a better assessment of the influence of the high temperature phase of mixtures of fatty alcohols and surfactants on the formation of lamellar gel networks. In particular, the force-field model presented here paves the road for extending MD simulations into multi-bilayer systems,  where one can test the influence of the bilayer-bilayer interactions on the build-up of the lamellar network .

\begin{acknowledgments}
T.E.O., F.L., L.N.M., A.L.F., B.Q., F.T., and C.M. acknowledge partial support from  the Investissements d'Avenir program ``D\'eveloppement de l'\'Economie Num\'erique'' through
the SMICE project. T.E.O., F.T., and C.M. acknowledge partial support from L'Or\'eal Research \& Innovation, and F.L., L.N.M., A.L.F., and B.Q. are employees of L'Or\'eal Research \& Innovation. We acknowledge support from the HPC cluster TGCC (Tr\`es Grand Centre de Calcul du CEA, Bruy\`eres-le-Ch\^atel, France).

\end{acknowledgments}

%

\end{document}